\documentclass[prd,onecolumn]{revtex4}
\begin{document}
\title{Modeling phantom energy wormholes from Shan-Chen fluids}
\author{Deng Wang}
\email{Cstar@mail.nankai.edu.cn}
\affiliation{Theoretical Physics Division, Chern Institute of Mathematics, Nankai University,
Tianjin 300071, China}
\author{Xin-he Meng}
\email{xhm@nankai.edu.cn}
\affiliation{{Department of Physics, Nankai University, Tianjin 300071, P.R.China}\\{
State Key Lab of Theoretical Physics, 
Institute of Theoretical Physics, CAS, Beijing 100080, P.R.China}}
\begin{abstract}
In recent years, cosmic observational data have reported that our present universe is undergoing an accelerated expansion, which has been termed as mysterious `` dark energy '' phenomena, that is, the origin of dark energy has not been determined yet. According to our previous work  that  a new equation of state \cite{X.Shan and H.Chen1993} can be employed to explain the dark energy, and we are very interested in investigating the astrophysical scale properties of dark energy based on the new cosmological fluids given that the universe is filled with a dark energy fluid everywhere. Hence, in this paper, we study the exact solutions of spherically-symmetrical Einstein field equations describing wormholes supported by phantom energy from Shan-Chen (SC) fluids  by considering an obvious relation between the transversal pressure and energy density which is different from our previous work \cite{111}. We have still investigated the important case $\psi\approx1$ which corresponds to the `` saturation effect '', and this regime corresponds to an effective form of `` asymptotic freedom '', occurring at cosmological rather than subnuclear scales. Then we find out two solutions by making some special choices for the shape function $b(r)$ and discuss the singularities of the solutions and find that the spacetimes are both geodesically incomplete. At the same time, it is worth noting that whether the solutions are geodesically complete or incomplete depends on the appropriate choice of the shape function $b(r)$ through our works. In addition, we match our interior solutions to the exterior Schwarzschild solutions and calculate out the total mass of the wormhole when $r\leq a$. Finally, we acquire that the surface stress-energy $\sigma$ is zero and the surface tangential pressure $\wp$ is positive when discussing the surface stresses of the solutions, and give out the perspectives on possible work in the future.
\end{abstract}
\maketitle
\section{introduction}
From the observational data of Supernova Type Ia (SN Ia) accumulated from the year 1998, Riess et al.\cite{Riess A.G.1998} in the High-redshift Supernova Search Team and Perlmutter et al. \cite{Perlmutter S.J.1999} in the Supernova Cosmology Project Team independently reported that the present universe expansion is accelerating. The source for the late-time cosmic acceleration was dubbed  as `` dark energy ''.  However, no theoretical model right determining the nature of dark energy is available at present. In recent years, there were a variety of models of dark energy proposed. For instance, the simplest candidate for dark energy is the so-called cosmological constant $\Lambda$, whose energy density remains constant. The effective $\Lambda$ (the quantity that appears in the Friedmann equations) is inferred from the cosmic observations when its theoretical value can be calculated accurately from quantum field theory. Comparing these quantities we find out $\Lambda^{th}/\Lambda^{obs}\sim 10^{72}GeV^4/10^{-47}GeV^4=10^{119}$ and it is not easy to realize the cosmological constant problem \cite{Weinberg}. Thus,
the actual origin of dark energy could not be $\Lambda$ (interpreting it as quantum vacuum). If so, one may seek for some alternative models to explain the cosmic acceleration phenomena today. Basically there are two approaches to construct models of dark energy. The first one is to modify the right-hand side (r.h.s.) of the Einstein field equations by considering specific forms of the energy-momentum tensor $T_{\mu\nu}$ with a negative pressure. The representative models that belong to this class are cosmon or quintessence model \cite{Y.Fujii,Ford,C.Wetterich,B.Ratra,S.M. Carroll,A.Hebecker,Turner,Caldwell} that make use of scalar fields with slowly varying potentials, or scalar field models with nonstandard kinetic terms (k-essence) \cite{C.Armendariz-Picon V.Mukhanov and P.J.Steinhardt 2000,T. Chi}, the Chaplygin gas model \cite{A.Y. Kamenshchik} and its generalizations \cite{M.C.Bento O.Bertlami and A.A.Sen 2002}, perfect fluid models \cite{A.Y. Kamenshchik, M.C.Bento O.Bertlami and A.A.Sen 2002} and dark fluid models \cite{rm}. The second approach for the construction of dark energy models is to modify the left-hand side (l.f.s.) of the Einstein field equations. The representative models that belong to this class (that we denote ``modified gravity'') are the so-called $f(R)$ gravity \cite{ S. Capozziello, S. Capozziello et al, S.M.}, braneworld models \cite{L.,Davli, V. Sahni} and cosmological models from scalar-tensor theories of gravity (see, e.g., Refs.\cite{L. Amendola,J.,T. Chiba,N.Bartolo,F.,V.Sahni and A.A.Starobinsky 2006, P.Ruiz-Lapuente2007} and references therein). Unfortunately, through some simple analysis, these models are all dominated by the samilar equation of state. Nevertheless, a new  modified equation of state has been first proposed by Shan and Chen (SC) in the context of lattice kinetic theory \cite{X.Shan and H.Chen1993}, with the primary intent that repulsion is replaced by a density-dependent attraction. Then, Donato Bini et al. have studied the dark energy phenomena from the cosmological fluids obeying a SC non-ideal equation of state \cite{Donato Bini et al2013}. The main idea is to postulate the cosmological fluids obey SC equation with `` asymptotic freedom ", namely, ideal gas behavior at both high and low density regimes, with a liquid-gas coexistence loop existence. Through some numerical calculations and analysis, they have found that a cosmological FRW fluid obeying the SC equation of state naturally evolves towards a present-day universe with a suitable dark-energy component, with no need of invoking any cosmological constant. To be precise, they have shown that, in the case of a simple model without any additional component in the cosmological fluid, starting from an ordinary equation of state at early times (e.g., satisfying the energy condition typical of a radiation-dominated universe), the SC pressure changes its sign at a certain time in the past and remains negative for a large time interval, including the present epoch.

 This is a nice idea to present a subtle explanation for the mysterious dark energy phenomena. Therefore, we are very interested in exploring the effects on astrophysics scales given that the dark energy is universal, that is, formed wormholes based on this new cosmological fluid if we assume the universe is permeated by a dark energy fluid. Moreover, our work is mainly motivated by three earlier studies \cite{Rahaman,P,Rahaman.} that have demonstrated the possible existence of wormholes in the outer regions of the galactic halo and in the central parts of the halo, respectively, based on NFW (Navarro-Frenk-White) density profile and the URC (Universal Rotation Curve) dark matter model \cite{R,NFW}. Especially, the second result is an important compliment to the earlier result, thereby confirming the possible existence of wormholes in most of the spiral galaxies. After their interesting and suggestive works, immediately, A. \"{O}vg\"{u}n et al. considered the existence of wormholes in the spherical stellar systems. And it is shown that based on the Einasto model \cite{J. Einasto,Einasto,D. Merritt} wormholes in the outer regions of spiral galaxies are possible while the central regions prohibit such space-time structure formations.

 Wormholes as kind of special spacetime structures connecting two different universes or two different regions of our own universe have been studied widely \cite{M.S.Moris and K.S.Thorne1988,Visser}. The earliest significant contribution we are aware of is the introduction, in 1935, of the object now referred to as Einstein-Rosen bridge \cite{Einstein}. This field then lay follow for twenty years until Wheeler first coined the term `` wormhole '' and introduced his idea of `` spacetime foam '' \cite{Wheeler}. A thirty year interregnum followed, punctuated by isolated contributions, until the major revival of interest following the 1988 paper by Morris and Thorne \cite{M.S.Moris and K.S.Thorne1988}. After that, the subject has grown substantially, and it is now almost out of control and there are many branches out of the special space-time structures\cite{M.S.Moris and K.S.Thorne1988,fc} etc, such as energy conditions, wormhole construction, stability, time machines, and astrophysical signatures.

 This imaginary and intuitive concept is one of the most popular and attractive area of research in general relativity. Recently, due to the new discovery that our universe is undergoing an accelerated expansion\cite{Riess A.G.1998,Perlmutter S.J.1999}, an increasing attention to these subjects has arisen significantly in connection with the discovery. Since in both cases the null energy conditions (i.e. NEC, which requires the stress-energy tensor $T_{\alpha\beta}k^\alpha k^\beta\geq0$ for all null vectors) are violated. Thus, an interesting and unexpected correlation between two seemingly separated subjects appeared. To be precisely, the accelerated expanding universe can be described by the Friedmann equation $\ddot{a}/a=-\frac{4\pi}{3}(\rho+3p)$, i.e. $\ddot{a}>0$. (Here we take units $G=c=1$). In papers \cite{mw,Peter K F kuhfitting 2006}, we have known the cosmic expansion acceleration is caused by a hypothetical negative pressure dark energy with the energy density $\rho>0$ and $p=-K\rho$ with $K>\frac{1}{3}$. The important situation of dark energy is phantom energy when $K>1$ and the same violation of the null energy condition occurs. In the quintessence models the parameter range is $1/3<K<1$ and the dark energy decreases with a scale factor $a(t)$ as $\rho_Q \propto a^{-3(1+\omega)}$, and case $K=1$ corresponds to a cosmological constant, and $K=2/3$ is extensively analyzed in work by\cite{P.F.Gonzales-Diaz 2002}.

One of the most important problems in general relativity is to find the exact solutions of Einstein field equations. In this paper, we will explore phantom energy wormholes from SC cosmological fluids with anisotropic pressure. For reference, see our previous work \cite{111}, two wormhole solutions with SC equation have been investigated. But, there is no sensible relation between a transversal pressure and the energy density. Our strategy which looks more physically reasonable is to define an equation of state and find the solutions afterwards.

The present paper is planned in the following manner. In next Section, we make a brief introduction to different kinds of cosmological model, phantom dark energy and SC equation of state. Moreover, we derive the field equation of the important case $\psi=1$ starting from a general line element and SC equation of state. At the same time, we consider an obvious relation between the transversal pressure and energy $p_\perp=\alpha(r)\rho$. In Section 3,  we make two special choices for the shape function: $b(r)=r_0+\frac{1}{k}(r-r_0)$ and $b(r)=r_0+\frac{1}{\sum_{n=1}^mA_nK^n}(r-r_0)$, respectively, and acquire the corresponding two solutions. Although they are neither asymptotically flat and geodesically complete, we work out the total mass of the wormholes when $r\leq a$ through matching the interior solutions of them to the exterior Schwarzschild solutions and find that the surface stress-energy $\sigma$ is zero and the surface tangential pressure $\wp$ is positive when discussing the surface stresses of the solutions. In final Section, we make some conclusions and discussions about our work and point out the possible work in the future.

\section{Wormholes as special space-time structures}
Considering the general line element
\begin{equation}
ds^2=-Udt^2+\frac{dr^2}{V}+r^2d\omega^2,d\omega^2=d\theta^2+sin^2\theta d\phi^2,V=1-\frac{b(r)}{r}.
\end{equation}

Here the radial coordinate $r$ runs in the range $r_0\leq r<\infty$. To describe a wormhole, the metric coefficient U should be finite and non-vanishing in the vicinity of $r_0$, and the shape function $b(r)$, which determine the spatial shape of the wormhole when viewed, must obey the usual flare-out conditions at the throat \cite{M.S.Moris and K.S.Thorne1988,Visser}:
\begin{equation}
b(r_0)=r_0,
\end{equation}
\begin{equation}
b'(r_0)<1,
\end{equation}
\begin{equation}
b(r)<r,r>r_0.
\end{equation}
Using Einstein field equation, $G_{\alpha\beta}=8\pi T_{\alpha\beta}$, in an orthonormal reference frame and considering $tt$ and $rr$ components, we have the following equations
\begin{equation}
\frac{b'}{8\pi r^2}=\rho(r),
\end{equation}
\begin{equation}
\frac{U'}{U}=\frac{8\pi p_r r^3+b}{r(r-b)},
\end{equation}
where $T_{t}^t=-\rho, T_{r}^r=p_r, T_{\nu}^\mu$ is the stress-energy tensor.

One can also derive from the conservation law of the stress-energy tensor $T^\nu_{\mu;\nu} = 0$ with $\mu=r$ that
\begin{equation}
p_\perp=\frac{r}{2}[p'_r+\frac{2p_r}{r}+\frac{U'}{2U}(p_r+\rho)],
\end{equation}
where $T_\theta^\theta=T_\phi^\phi=p_\perp$. From now on we assume that our source of gravitation is realized by the phantom energy with the SC equation of state that contains a radial pressure [8]
\begin{equation}
p_r=-K\rho_{(crit),0}[\frac{\rho}{\rho_{(crit),0}}+\frac{g}{2}\psi^2],
\end{equation}
\begin{equation}
\psi=1-e^{-\beta\frac{\rho}{\rho_{(crit),0}}}.
\end{equation}
Where $\rho_{(crit),0}=3(H_0)^2/8\pi$ is the present value of the critical density ($H_0$ denoting the Hubble constant) and the dimensionless quantities $K$, $g\leq0$ and $\beta\geq0$ can be regarded as free parameters of the model.

Since we find two solutions when $\rho\gg\rho_\ast$, $\rho_\ast=\rho_{(crit),0)}/\beta$ ($\rho_\ast$ being the typical density above which $\psi$ undergoes a `` saturation effect "), i.e., $\psi\approx1$ \cite{Donato Bini et al2013}, this regime corresponds to an effective form of asymptotic freedom. So the SC equation of state (10) becomes
\begin{equation}
p_r=-K\rho_{(crit),0}[\frac{\rho}{\rho_{(crit),0}}+\frac{g}{2}].
\end{equation}
Since the pressure $p_r$ is negative, so $-\rho<\frac{g}{2}\rho_{(crit),0}<0$ and $K>\frac{\rho}{\rho+\frac{g}{2}\rho_{(crit),0}}>1$ corresponds to a SC's version of phantom energy.

At the same time, we supposed also that the pressures are anisotropic and
\begin{equation}
p_\perp=\alpha(r)\rho.
\end{equation}
Hence, we consider a simple relation between the tangential pressure and energy density but with $p_\perp\neq p_r$.
\section{exact solutions}
\subsection{Special choice for the shape function: $b(r)=r_0+\frac{1}{k}(r-r_0)$}
First of all, we choose $b'(r)=1/K$ which is the solution in [9], so
\begin{equation}
b(r)=r_0+\frac{1}{k}(r-r_0),
\end{equation}
and naturally the energy density is
\begin{equation}
\rho(r)=\frac{1}{8\pi r^2K}.
\end{equation}
Substitution of Eqs.(10) and (12) into Eq.(6) leads to
\begin{equation}
\frac{U'}{U}=-\frac{1}{r}-\frac{4\pi Kg\rho_{(crit),0}r^2}{(r-r_0)(1-\frac{1}{K})}.
\end{equation}
This equation is readily solved and leads to
\begin{equation}
U(r)=\frac{C}{r}e^{\frac{-2gK^2\pi\rho_{(crit),0}[r(2r_0+r)+2r_0^2\ln(r-r_0)]}{K-1}}.
\end{equation}
where $C$ is an arbitrary constant that affects the normalization of time. So the line element can be written as
\begin{equation}
ds^2=-\frac{C}{r}e^{\frac{-2gK^2\pi\rho_{(crit),0}[r(2r_0+r)+2r_0^2\ln(r-r_0)]}{K-1}}dt^2+\frac{1}{(1-\frac{1}{K})(1-\frac{r_0}{r})}dr^2+r^2(d\theta^2+sin^2\theta d\phi^2).
\end{equation}
  It is easy to check that the shape function $b(r)$ satisfies the usual flare-out conditions Eqs.(2-4) and $V(r)=(1-\frac{1}{K})(1-\frac{r_0}{r})$ in our consideration here. In addition, the $tt$ and $rr$ components of the solution respectively go to zero and infinity when $r$ goes to $r_0$, respectively. So an event horizon is located at $r=r_0$, implying the existence of a non-traversable wormhole. At the same time, obviously, the solution is geodesically incomplete (we can also see the r.h.s. of Eq.(14) will go to infinity when $r=r_0$).

Then our next step is to work out the function $\alpha(r)$ in $p_\perp$ corresponding to the above metric. Substitute Eqs.(10-11) and (13-14) into Eq.(7), we have the following equation:
\begin{equation}
\alpha(r)=\frac{K-1}{4}-3\pi K^2g\rho_{(crit),0}r^2+\frac{\pi K^2g\rho_{(crit),0}r^3}{r-r_0}-\frac{4\pi^2K^4g^2\rho_{(crit),0}^2r^5}{(r-r_0)(1-K)}.
\end{equation}
We can find the first term $\frac{K-1}{4}$ is the value of $\alpha$ in the paper [8], and the last three terms occur since we have used the SC equation of state. By some simple analysis, one can realize that the second and the last terms are both positive and the third is negative as well as the last term play a more role when $r$ become more and more larger.

  Moreover, it is easily to be seen that the metric is not asymptotically flat, however, it can be glued to external Schwarzschild solution at some $r$:
\begin{equation}
ds^2=-(1-\frac{2M}{r})dt^2+(1-\frac{2M}{r})dr^2+r^2(d\theta^2+\sin^2\theta d\theta^2).
\end{equation}

To match the interior to the exterior, one needs to apply the junction conditions that follow the theory of general relativity. If there is no surface stress-energy  terms at the surface S, the junction is called a boundary surface. If, on the other hand, surface stress-energy terms are present, the junction is called a thin-shell \cite{Lobo F S N 2005}.

 A wormhole with finite dimensions, in which the matter distribution extends from the throat, $r=r_0$, to a finite distance $r=a$, obeys the condition that the metric is continuous. Due to the spherical symmetry the components $g_{\theta\theta}$ and $g_{\phi\phi}$ are already continuous, so one needs to impose continuity only on the remaining components at $r=a$:
\begin{equation}
g_{tt(int)}(a)=g_{tt(ext)}(a),
\end{equation}
\begin{equation}
g_{rr(int)}(a)=g_{rr(ext)}(a).
\end{equation}
Equations (18) and (19) are for the interior and exterior components, respectively. These requirements, in turn, lead to
\begin{equation}
\Phi_{int}(a)=\Phi_{ext}(a),
\end{equation}
\begin{equation}
b_{int}(a)=b_{ext}(a).
\end{equation}
Particularly,
\begin{equation}
e^{2\alpha(r)}=\frac{1}{1-\frac{b(a)}{a}}=\frac{1}{1-\frac{2M}{a}}.
\end{equation}

So, one can deduce the mass of the wormhole is given by
\begin{equation}
M=\frac{b(a)}{2}=\frac{1}{16\pi a^2K}.
\end{equation}
According to Eqs. (16) and (18-19), we now have
\begin{equation}
C=(a-\frac{1}{8\pi a^2K})e^{\frac{2gK^2\pi\rho_{(crit),0}[a(2r_0+a)+2r_0^2\ln(a-r_0)]}{K-1}}.
\end{equation}
Therefore, the line element becomes
\begin{equation}
ds^2=-\frac{1}{r}(a-\frac{1}{8\pi a^2K})e^{\frac{2gK^2\pi\rho_{(crit),0}[a(2r_0+a)-r(2r_0+r)+2r_0^2\ln(\frac{a-r_0}{r-r_0})]}{K-1}}dt^2+\frac{1}{(1-\frac{1}{K})(1-\frac{r_0}{r})}dr^2+r^2(d\theta^2+sin^2\theta d\phi^2).
\end{equation}
Our next step is to take into account the surface stresses. Using the Daromis-Israel formalism \cite{N.Sen 1924,K.Lanczos 1924}, the surface stresses are given by
\begin{equation}
\sigma=-\frac{1}{4\pi a}(\sqrt{1-\frac{2M}{a}}-\sqrt{1-\frac{b(a)}{a}}),
\end{equation}
\begin{equation}
\wp=\frac{1}{8\pi a}(\frac{1-\frac{M}{a}}{\sqrt{1-\frac{2M}{a}}}-[1+a\Phi'(a)]\sqrt{1-\frac{b(a)}{a}}).
\end{equation}

It is so clear that the surface stress-energy $\sigma$ is zero and the surface tangential pressure $\wp$ is positive.\\
Subsequently, another important consideration is singularity of the solution. To see this, we can deduce that $V(r)=0$ when $r=r_0$, and it's actually a physical singularity. Then let us consider a free particle with an energy $E=-u_0$ and angular momentum $L=u_{\phi}$ (in dimensionless units) moving in the Eq.(16), $u^{\mu}$ being the 4-velocity. Usually, we choose the plane to be $\theta=\pi/2$, so one can easily obtain from the condition $g_{\mu\nu}u^{\mu}u^{\nu}=\varepsilon$ ($\varepsilon=0$ for massless particles and $\varepsilon=-1$ for massive particles) that
\begin{equation}
(u^r)^2=V(r)(\frac{E^2}{-e^{2C}}-\frac{L^2}{r^2}+\varepsilon).
\end{equation}
Here, for simplicity, we just consider radial null geodesics ($L=0, \varepsilon=0$) and it follows that
\begin{equation}
(\frac{dr}{d\tau})^2=\frac{1-K}{CK}\frac{r-r_0}{e^{\frac{-2gK^2\pi\rho_{(crit),0}[r(2r_0+r)+2r_0^2\ln(r-r_0)]}{K-1}}}.
\end{equation}
Obviously, the numerator and denominator will both approach 0 when $r=r_0$. Thus, through mathematically so-called `` L'Hospital rule '', we can get
\begin{equation}
(\frac{dr}{d\tau})^2=\frac{1-K}{CK}\frac{1+\frac{4gK^2\pi\rho_{(crit),0}r_0^2}{K-1}}{e^{\frac{-2gK^2\pi\rho_{(crit),0}[r(2r_0+r)+2r_0^2\ln(r-r_0)]}{K-1}}},
\end{equation}
it diverges when $r=r_0$ very clearly. Therefore, the solution is geodesically incomplete.
\subsection{Special choice for the shape function: $b(r)=r_0+\frac{1}{\sum_{n=1}^mA_nK^n}(r-r_0)$}
Another possibility is choosing $b'(r)=\frac{1}{\sum_{n=1}^mA_nK^n}$, so
\begin{equation}
b(r)=r_0+\frac{1}{\sum_{n=1}^mA_nK^n}(r-r_0),
\end{equation}
where $A_n$ is a set of arbitrary positive integers and $m$ is also a limited positive integer. And the energy density is
\begin{equation}
\rho(r)=\frac{1}{8\pi r^2\sum_{n=1}^mA_nK^n}.
\end{equation}
Here we still consider the important case $\psi\approx1$, since the quantity $\psi$ can be interpreted as the density of a chameleon scalar field \cite{J.Khoury and A.Weltman 2004} which is used for reconciling large coupling models with local gravity constraints. Substitution of Eqs.(10) and (18) into Eq.(6) leads to
\begin{equation}
\frac{U'}{U}=-\frac{1}{r}-\frac{4\pi Kg\rho_{(crit),0}r^2}{(r-r_0)(1-\frac{1}{\sum_{n=1}^mA_nK^n})}.
\end{equation}
It follows that
\begin{equation}
U(r)=\frac{C'}{r}e^{\frac{-2g\sum_{n=1}^mA_nK^{n+1}\pi\rho_{(crit),0}[r(2r_0+r)+2r_0^2\ln(r-r_0)]}{\sum_{n=1}^mA_nK^n-1}},
\end{equation}
where $C'$ is another arbitrary constant that affects the normalization of time. Now the line element is
\begin{equation}
ds^2=-\frac{C'}{r}e^{\frac{-2g\sum_{n=1}^mA_nK^{n+1}\pi\rho_{(crit),0}[r(2r_0+r)+2r_0^2\ln(r-r_0)]}{\sum_{n=1}^mA_nK^n-1}}dt^2+\frac{1}{(1-\frac{1}{\sum_{n=1}^mA_nK^n})(1-\frac{r_0}{r})}dr^2+r^2(d\theta^2+sin^2\theta d\phi^2).
\end{equation}
Obviously, the shape function $b(r)$ also satisfies the flare-out conditions here. Moreover, if $m=1$ and $A_n=1$, the solution will reduce to Eq.(16).

Then the function $\alpha(r)$ here is obtained by substituting Eqs.(13-14) and (18-19) into Eq.(7)
\begin{equation}
\alpha(r)=\frac{K-1}{4}-3\pi g\sum_{n=1}^mA_nK^{n+1}\rho_{(crit),0}r^2-\frac{\pi Kg\rho_{(crit),0}r^3(1-K)}{(r-r_0)(1-\frac{1}{\sum_{n=1}^mA_nK^n})}+\frac{4\pi^2K^2g^2\rho_{(crit),0}^2r^5}{(r-r_0)(\sum_{n=1}^mA_nK^n-1)}.
\end{equation}
We also can find the first term $\frac{K-1}{4}$ is the value of $\alpha$ in the paper [8], and the last three terms occur since we have used the SC equation of state. Similarly, from Eq.(31) we can see the metric is geodesically incomplete since there is an event horizon when $r=r_0$ and the wormhole is non-traversable. One can also get this conclusion by the same method as the first solution. At the same time, we can find that it's also not asymptotically flat and we will also construct the asymptotically flat region through gluing it to the external Schwarzschild solution at some $r=a$. So we can deduce the mass of this wormhole:
\begin{equation}
M=\frac{b(a)}{2}=\frac{1}{16\pi a^2\sum_{n=1}^mA_nK^n}.
\end{equation}
Then according to Eqs.(18-19) and (31), we now have
\begin{equation}
C'=(a-\frac{1}{8\pi a^2\sum_{n=1}^mA_nK^n})e^{\frac{2g\sum_{n=1}^mA_nK^{n+1}\pi\rho_{(crit),0}[a(2r_0+a)+2r_0^2\ln(a-r_0)]}{\sum_{n=1}^mA_nK^n-1}}.
\end{equation}
So the metric becomes
\begin{equation}
ds^2=-\frac{1}{r}e^{\frac{2g\sum_{n=1}^mA_nK^{n+1}\pi\rho_{(crit),0}[a(2r_0+a)-r(2r_0+r)+2r_0^2\ln(\frac{a-r_0}{r-r_0})]}{\sum_{n=1}^mA_nK^n-1}}dt^2+\frac{1}{(1-\frac{1}{\sum_{n=1}^mA_nK^n})(1-\frac{r_0}{r})}dr^2+r^2(d\theta^2+sin^2\theta d\phi^2).
\end{equation}
Then take into account the the surface stresses in the forms of Eqs.(27-28) again, the same conclusion that the surface stress-energy $\sigma$ is zero and the surface tangential pressure $\wp$ is positive is obtained. Thus, one can realize that we have found out a family of static spherically-symmetrical solutions of the Einstein field equations by a appropriate choice of $b(r)=r_0+\frac{1}{\sum_{n=1}^mA_nK^n}(r-r_0)$ which will reduce to the first solution when $m=1$ and $A_n=1$.

\section{Conclusions and discussions}
  In summary, we have three main reasons to investigate the wormhole solutions based on the SC cosmological fluids. First, the explanation of the cosmic acceleration expansion requires the introduction of either cosmological constant, or of a mysterious dark energy (time dependent or extended gravities), filling the universe and dominating its expansionary evolution. Given that the universe is permeated by a dark energy fluid, therefore, we should investigate the astrophysical scale properties of dark energy. Secondly, the NEC energy condition is both violated in our accelerating universe phenomena and the possible existence of wormholes,  and the unexpected overlap between two separated subjects is very exciting. On the one hand, phantom energy is one important kind of dark energy model case when parameter $K>\frac{\rho}{\rho+\frac{g}{2}\rho_{(crit),0}}>1$ in the SC equation of state. On the other hand, the violation of the null energy condition of dark energy naturally leads to the occurrence of phantom energy that is always called "exotic matter", which is usually to be satisfied with the existence of wormholes in the literature. Thirdly, three early studies \cite{Rahaman,P,Rahaman.} give us strong enough motivation to investigate wormholes by using different density profile. Moreover, in our entire universe, existence of wormholes is an important problem in physics both at micro and macro scales. There is no doubt that together with black holes, pulsars (physical neutron stars) and white darfs, etc, wormholes constitute the most interesting, strange and puzzling astrophysics objects. Therefore, in this paper, we further investigate the exact solutions of spherically-symmetrical Einstein field equations describing wormholes supported by phantom energy from SC fluids,  by considering an obvious relation between the transversal pressure and energy density which is different from our previous work results\cite{111}. It is shown that the investigated important case $\psi\approx1$ in SC model which corresponds to the `` saturation effect '', corresponds to an effective form of `` asymptotic freedom '' regime, occurring at cosmological rather than subnuclear scales. Whence we find out two simple solutions of spherically-symmetrical Einstein equations with SC equation of state describing a wormhole. Then we carefully explore the two solutions by making some special choices for the shape function $b(r)$. At the same time, we discuss the singularities of the solutions and find that the spacetimes are both geodesically incomplete. And it is worth noting that whether the solutions are geodesically complete or incomplete depends on the appropriate choice of the shape function $b(r)$, through our works. In addition, we match our interior solutions to the exterior Schwarzschild solutions and calculate out the total mass of the wormhole when $r\leq a$. Finally, we acquire that the surface stress-energy $\sigma$ is zero and the surface tangential pressure $\wp$ is positive when discussing the surface stresses of the solutions.
  
  The century celebrating general relativity though very successful when testing from sub-mm to the solar system scales, is still full of implying puzzlings, especially when confronting with the dark matter and dark energy mysteries in various scales if the dark sectors are universal existing.
  Wormholes are theoretically special objects in Universe space-time constructions which now appear to attract more observational astrophysics efforts and may provide a new window  for new physics discovery.
  It would be desirable to find out traversable wormholes by making some subtle choices for the shape function or the redshift function, since phantom traversable wormholes have far-reaching physical and cosmological implications. For an instance, an advanced civilization may use these special geometries to induce closed timelike curves, consequently violating causality and constructing the time-machines. And apply the SC equation of state to the modified theories of gravity may also be more interesting such as $f(R)$ gravity, although that will be more complicated.

\begin{center}
\textbf{Acknowledgments}
\end{center}

For helpful discussions and comments, we thank Prof. Saibal Ray and Jingling Chen, Qixiang Zou, Guang Yang, Shengsen Lu, and Dr F. Canfora for useful communications. This work is in part supported by the National Science Foundation of China.

\end{document}